\documentclass[10pt,aps,prb,twocolumn,groupedaddress,showpacs,showkeys]{revtex4-1}
\usepackage{amsmath}
\usepackage{braket}
\usepackage[]{graphicx}
\usepackage{bm}
\usepackage{booktabs}
\usepackage{makecell}
\usepackage{multirow}
\newlength{\figwidth} 
\newcommand{\beq}{\begin{equation}}

\newcommand{\beql}[1]{\begin{equation}\label{#1}}
\newcommand{\eeq}{\end{equation}}
\newcommand{\bsp}{\begin{split}}
\newcommand{\esp}{\end{split}}

\newcommand{\Eq}[1]{Eq.~(\ref{#1})}
\newcommand{\Equation}[1]{Equation~(\ref{#1})}

\newcommand{\Fig}[1]{Fig.~\ref{#1}}

\newcommand{\Figure}[1]{Figure~\ref{#1}}
\newcommand{\Table}[1]{Table.~(\ref{#1})}

\begin{document}

\title{Generalized coupled mode formalism in reciprocal waveguides with gain/loss, anisotropy or bianisotropy}
\author{Weijin Chen} 
 \affiliation{School of Optical and Electronic Information, Huazhong University of Science and Technology, Wuhan 430074,  China.}
\author{Zhongfei Xiong}
 \affiliation{School of Optical and Electronic Information, Huazhong University of Science and Technology, Wuhan 430074,  China.}
\author{Jing Xu}
\affiliation{School of Optical and Electronic Information, Huazhong University of Science and Technology, Wuhan 430074,  China.}
\author{Yuntian Chen} \email{yuntian@hust.edu.cn}
\affiliation{School of Optical and Electronic Information, Huazhong University of Science and Technology, Wuhan 430074,  China.}

\date{\today}
\begin{abstract}
In  anisotropic or bianisotropic waveguides, the standard coupled mode theory fails due to the broken link between the forward and backward propagating modes, which together form the dual mode sets that are crucial in constructing  couple mode equations. We generalize the coupled mode theory by treating  the forward  and backward propagating   modes on the same footing via a generalized eigenvalue problem that is exactly equivalent to the  waveguide Hamiltonian.  The generalized eigenvalue  problem is fully characterized by two operators, i.e., $( \bar{\bm{L}},\bar{\bm{B}})$, wherein $\bar{\bm{L}}$ is a self-adjoint differential operator, while $\bar{\bm{B}}$ is a constant antisymmetric operator. From the properties of $\bar{\bm{L}}$ and $\bar{\bm{B}}$, we establish the relation between the dual mode sets that are essential in constructing coupled mode equations in terms of forward and backward propagating modes. By perturbation, the generalized coupled mode equation can be derived in a natural way.  Our generalized coupled mode formalism can be used to study the mode coupling in  waveguides that may contain  gain/loss, anisotropy or bianisotropy.  We further illustrate how the generalized coupled theory can be used to study the modal coupling in anisotropy and bianisotropy waveguides through a few concrete examples.

\end{abstract}

\pacs{0000}

\maketitle

\section{Introduction}

Coupled mode theory (CMT)  is an indispensable tool to analyze and design the photonic devices, such as  waveguides and cavities\cite{Kocabaş2009pr,Xu2000pr,Paddon2000pr,Snyder1972,Yariv1973,Hardy1985,Huang1994,Chak2006pr,Villeneuve1996pr,Wubingbing2016,Fan1999,Haus1984}, and has far-reaching implications and applications in many subfields of  optics\cite{Sivan2016pr,Bowen2014pr,Hamam2007pr,Chak2006pr,Chak2006pr,Buschlinger2017pr,Villeneuve1996pr,Kocabaş2009pr,Xu2000pr,Ruan2012pr,Paddon2000pr,Malomed2005pr,Carnevale1982pr,Sun2011pr,Iizuka2000pr,Qiu2009,Johnson2002}. CMT is a  theoretical framework that treats each individual optical mode with certain spatiotemporal distribution as a single object, among which one mode could be coupled to  another  as the control parameter of the optical  system, such as refractive indices or shapes of optical structures, varies.  As a simple mode, CMT   provides  not only an intuitive  picture of how the photonic modes are hybridized, but also provides quantitative assessment of how the hybridization among those relevant modes evolves. Especially, CMT in the waveguides has been a great tool to study the modal properties of  various waveguides\cite{Kocabaş2009pr,Paddon2000pr,Yariv1999} as well as waveguide lattices\cite{Peng2012pr,Malomed2005pr}. In the coupled waveguide-cavity systems\cite{Xu2000pr,Waks2005,Kristensen2017}, the temporal CMT\cite{Lifante2003Chapter4} provides insight to design the ultrafast optical switch. Considering the recent development in metamaterial, man-made anisotropic medium can be created, leading to interesting applications in controlling the flow and polarization of light\cite{Lalanne2008}. Therefore, there is a large need to study  waveguides that may contain  anisotropy or bianisotropy.

% Here we show a figure to illustrator what kind of the waveguide, bianisotropic or anisotropic waveguides, with the help of figures. 

In waveguide CMT, each mode charactered by the propagation constant $\beta$ and the corresponding mode profile  can be treated as a single object described as follows,
\beq\label{ham}
\mathcal{H} {\bm \phi}=\beta {\bm \phi}
\eeq 
where $\mathcal{H}$ is the waveguide Hamiltonian. The construction of coupled mode equation of waveguides under certain perturbation    $\mathcal{H}+\Delta$ takes three  steps. Firstly, the field of the perturbed waveguide is expanded as ${\bm \phi}^{new}=\sum a_i {\bm \phi}_i$ where all ${\bm \phi}_i$ functions  span a complete mode set associated with Hamiltonian $\mathcal{H}$, i.e., the space of the right eigenstates $\left[{\bm \phi}_i\right]$; Secondly, \Eq{ham} of the perturbed system can be revised as a residual form $R=(\mathcal{H} +\Delta){\bm \phi}^{new}-\beta {\bm \phi}^{new}$, which is further tested against all the possible ${\bm \psi}_j$  coined as the test functions that span the space of the left eigenstates $\left[{\bm \psi}_j\right]$ associated with Hamiltonian $\mathcal{H}$. Lastly, the link between the expansion functions ${\bm \phi}_i$ and the test functions ${\bm \psi}_j$ can be built within a proper inner product between ${\bm \phi}_i$ and ${\bm \psi}_j$, as such the coupled mode equations can be constructed. The aforementioned procedure in constructing CMT has been used extensively in computational electromagnetism, such as method of moments \cite{Harrington1994MOM} and finite element method \cite{JinFem2002},  which can be further proved to be exactly equivalent to variational principle \cite{HausJLT1987,JinFem2002,Harrington2001,Pintus2014,Stakgold1968,Cvetkovic1986,Ruey-Beei Wu1986}.  

There are two different types of inner product, i.e., complex inner product and scalar inner product\cite{Rumsey1954,Chew2008}, both of which are heavily used in waveguide CMT. In the scheme of complex inner product, it turns out that the left eigenstates ${\bm \psi}_j$ can be obtained by performing the Hermitian operation on the right eigenstates ${\bm \phi}_i$, implying the fact that the waveguide Hamiltonian $\mathcal{H}$ is Hermitian operator and the integrated power flux flowing along propagation direction, i.e., $z$-axis, of the waveguide is conserved \cite{HausJLT1987,Marcuse1975,ChuangJLT1987}. In parallel, the left eigenstates ${\bm \psi}_j$ in a scalar inner product  can be obtained by performing the transpose operation  and certain operations in field components of the right eigenstates ${\bm \phi}_i$,  provided that the waveguide medium is reciprocal and the reaction conservation is fulfilled \cite{Xu2015,ChuangJLT1987}. In the waveguide with gain/losses, the integrated power flux flowing along the waveguide is apparently not conserved, thus  the modal coupling model based on the scalar inner product is proposed by Xu et. al. \cite{Xu2015} to study $\mathcal{PT}$-symmetric waveguides that contain balanced gain and losses. Essentially, the GCMT model \cite{Xu2015} is a revision of CMT by Haus \cite{HausJLT1987} by replacing  the complex inner product with the scalar inner product, wherein the reciprocity or the reaction conservation still holds.

Physically, the expansion function ${\bm \phi}_i$ can be taken as the forward propagating modes associated with the waveguide Hamiltonian $\mathcal{H}$, while the test function ${\bm \psi}_j$  corresponds to the time reversal partner (backward propagating modes) with respect to ${\bm \phi}_i$ in the scheme of complex (scalar) inner product. In either complex or scalar inner product, the inner product between ${\bm \phi}_i$ and ${\bm \psi}_j$ yields a physical quantity that is independent of $z$-coordinate, as such the three-dimensional (3D) waveguide problem is reduced to a 2D counterpart. Such dimensional reduction is necessary, which significantly reduces the computational load. In the scalar inner product, such dimensional reduction in CMT works perfect well for waveguide, in which the material is isotropic or only contains the in-plane anisotropy. In contrast, dimensional reduction is in conflict with the definite relation between the forward propagating modes ${\bm \phi}_i$ with the backward propagating modes ${\bm \psi}_j$, if the material tensor of waveguide contains terms  that couple the traverse  components and longitudinal component. In generic scenario,  the definite  relation between forward and backward propagating modes is lost. However, the coupled mode equations in all the available CMT in  literatures essentially  are constructed  by testing \Eq{ham} against all the possible ${\bm \psi}_j$, which is deduced from the expansion functions ${\bm \phi}_i$ \citep{HausJLT1987,Xu2015,JinFem2002,ChuangJLT1987,Harrington1994MOM} based on the aforementioned relation. Importantly, it is the completeness of  the mode set $\left[{\bm \psi}_i\right]$  of test functions, i.e., all the possible ${\bm \psi}_j$, that guarantees the equivalence to the variational principle. In this regard, the completeness  of $\left[{\bm \psi}_i\right]$  associated with the Hamiltonian $\mathcal{H}$ of the anisotropic/bianisotropic waveguides need to be restored to construct the coupled mode equations.

In special cases, the forward and backward propagating modes are related by symmetry operations \cite{xzf_arxiv}. In a generic waveguide with bianisotropy, the relation between the forward and backward propagating modes is lost, which indicates that there is no definite relation between the expansion function ${\bm \phi}_j$ and the  test function  ${\bm \psi}_j$. Hence, the procedures in constructing the coupled mode equations as given in literatures  \cite{HausJLT1987} do not apply here.  In this work, we start with a formal description of generalized eigenvalue problem of the bianisotropic waveguide by putting the forward and backward propagating modes on the same footing. Namely, the forward and backward  propagating modes are combined together to constitute the complete expansion mode set $\left[{\bm \phi}_i\right]$. As for the reciprocal waveguide, it  can be proved that the test function mode set $\left[{\bm \psi}_i\right]$  and expansion mode set $\left[{\bm \phi}_i\right]$ can be related. As such, the coupled mode equation can be constructed in a similar procedure as discussed  previously.

This paper is organized as follows. In Sec. II, we reformulate the vectorial wave equation of the waveguide problem into a generalized eigenvalue problem $(\bar {\bm L}, \bar{ \bm B})$. We further introduce the adjoint generalized eigenvalue problem $(\bar{\bm L^a}, \bar{\bm B^a})$ under the scalar inner product. In Sec. III, we examine the relation between the two dual waveguide systems, i.e., $(\bar{\bm L}, \bar{\bm B})$ and $(\bar{\bm L^a}, \bar{\bm B^a})$, in reciprocal waveguides. The symmetric relation between the modes associated with  $(\bar{\bm L}, \bar{\bm B})$ and $(\bar{\bm L^a}, \bar{\bm B^a})$ can be summarized by the 
\textit{same-$\beta$ argument} and the \textit{paring-$\beta$ argument}. Based on the symmetric relations, we provided the procedures of constructing CMT, in which the forward and backward propagating modes are included in the modal expansion set.  In Sec. IV,  we apply our theory to study three examples, showing that our theory captures the features of the modal coupling  due to broken link between the forward and backward propagating modes. Section V concludes the paper.

\section{Generalized eigenvalue problem of waveguide}

\subsection{Generalized eigenvalue problem of original waveguide system}
We consider a generic bianisotropic waveguide, in which the constitutive relation can be given as follows \cite{bian}:
\beq
\label{DB}
\begin{split}
\bm D &={\bar {\bm\varepsilon}} \bm E + {\bar{\bm \chi } }_{eh} \bm H, \\
\bm B &={\bar {\bm\mu}}\bm H + {\bar{\bm \chi } }_{he} \bm E,
\end{split}
\eeq
where $\bm D$/$\bm B$ is  electric displacement/magnetic induction, $\bm E$/$\bm H$  is electric/magnetic field respectively, ${\bar {\bm\varepsilon}}=\varepsilon_0 {\bar {\bm\varepsilon}}_r$ ($\bar {\bm\mu}=\mu_0 {\bar {\bm\mu}}_r$) is permittivity (permeability) tensor, ${\bar{\bm \chi } }_{eh}=\sqrt{\varepsilon_0 \mu_0}{\bar{\bm \chi } }_{r,eh}$ and ${\bar{\bm \chi } }_{he}=\sqrt{\varepsilon_0 \mu_0}{\bar{\bm \chi } }_{r,he}$ are magnetoelectric coupling constants. Explicitly, 
$\bar{\bm\varepsilon}_r=\begin{pmatrix}
 \bar{\bm\varepsilon}_r^{tt}&\bar{\bm\varepsilon}_r^{tz}\\
\bar{\bm\varepsilon}_r^{zt}&\bar{\bm\varepsilon}_r^{zz} \\
\end{pmatrix}=\begin{pmatrix}
 \varepsilon_r^{xx}&\varepsilon_r^{xy}&\varepsilon_r^{xz}\\
\varepsilon_r^{yx}&\varepsilon_r^{yy}&\varepsilon_r^{yz} \\
\varepsilon_r^{zx}&\varepsilon_r^{zy}&\varepsilon_r^{zz}\\
\end{pmatrix}$, $\bar{\bm\mu}_r=\begin{pmatrix}
 \bar{\bm\mu}_r^{tt}&\bar{\bm\mu}_r^{tz}\\
\bar{\bm\mu}_r^{zt}&\bar{\bm\mu}_r^{zz} \\
\end{pmatrix}=\begin{pmatrix}
 \mu_r^{xx}&\mu_r^{xy}&\mu_r^{xz}\\
\mu_r^{yx}&\mu_r^{yy}&\mu_r^{yz} \\
\mu_r^{zx}&\mu_r^{zy}&\mu_r^{zz}\\
\end{pmatrix}$, $\bar{\bm \chi} _{r,he}=
\begin{pmatrix}
 \bar{\bm\chi}_{r,he}^{tt}&\bar{\bm\chi}_{r,he}^{tz}\\
\bar{\bm\chi}_{r,he}^{zt}&\bar{\bm\chi}_{r,he}^{zz} \\
\end{pmatrix}=
i \left({\begin{array}{*{20}{c}}
  \chi^{xx}_{r,he}&\chi^{xy}_{r,he}&0 \\ 
  \chi^{yx}_{r,he}&\chi^{yy}_{r,he}&0 \\
  0&0&0\\
\end{array}} \right)$, and 
 $\bar{\bm \chi} _{r,eh}=\begin{pmatrix}
 \bar{\bm\chi}_{r,eh}^{tt}&\bar{\bm\chi}_{r,eh}^{tz}\\
\bar{\bm\chi}_{r,eh}^{zt}&\bar{\bm\chi}_{r,eh}^{zz} \\
\end{pmatrix}=i\left({\begin{array}{*{20}{c}}
  \chi^{xx}_{r,eh}&\chi^{xy}_{r,eh}&0 \\ 
  \chi^{yx}_{r,eh}&\chi^{yy}_{r,eh}&0 \\
  0&0&0\\
\end{array}} \right)$. 
Considering an infinitely long waveguide with translation symmetry along $z$, the waveguide modes can be obtained by solving  the Maxwell's equation with  a time harmonic dependence $e^{i\omega t}$, i.e., $
\nabla \times \bm E =-i\omega\left(\bar{\bm \mu}\bm H+{\bar{\bm\chi}}_{he}\bm E\right)$,
$\nabla \times \bm H=i\omega\left(\bar{\bm \varepsilon}\bm E+{\bar{\bm\chi}}_{eh}\bm H\right)$, 
where $i=\sqrt{-1}$. With normalization, i.e., $\bm e=\bm E$ and $\bm h=\sqrt{\frac{\mu_0}{\varepsilon_0}} \bm H$, we reformulate Maxwell's equations as follows, 
\begin{equation}\label{max1}
\begin{array}{c}
 \left[ {\nabla  \times  +i{k_0}{{\bar{\bm \chi } }_{he}^r}} \right]{\bm e_{3d}}(x,y,z){\text{ + }}i{k_0}{{\bar {\bm\mu} }_r}{\bm h_{3d}}(x,y,z) = 0, \\
  \left[ {\nabla  \times  - i{k_0}{{\bar{\bm \chi } }_{eh}^r}} \right]{\bm h_{3d}}(x,y,z) - i{k_0}{{\bar {\bm\varepsilon} }_r}{\bm e_{3d}}(x,y,z) = 0, 
\end{array}
\end{equation} 
where vacuum wavenumber $k_0=\omega \sqrt{\varepsilon_0 \mu_0}$. By translation symmetry alone $z$, the normalized electromagnetic field can be separated as the transverse terms and the longitudinal term, i.e., $ {\bm e_{3d}}(x,y,z) = {e_{2d}}(x,y){e^{ - i\beta z}} = \bm e_{2d}^t(x,y){e^{ - i\beta z}} + e_{2d}^z(x,y){e^{ - i\beta z}}$ (${\bm h_{3d}}(x,y,z) = {\bm h_{2d}}(x,y){e^{ - i\beta z}} = \bm h_{2d}^t(x,y){e^{ - i\beta z}} + h_{2d}^z(x,y){e^{ - i\beta z}}$), where $\beta$ is the propagation constant. Subsequently, \Eq{max1} can further be reformulated into 4 components equation by eliminating the longitudinal terms $e^z_{2d}(x,y)$ and $h^z_{2d}(x,y)$ via the expressions ${e_{2d}^z(x,y) = \frac{{{\nabla _t} \times h_{2d}^t(x,y) - i{k_0}\bar{ \bm \epsilon} _r^{zt} \cdot e_{2d}^t(x,y)}}{{i{k_0}\bar{ \bm \epsilon} _r^{zz}}}}$ and ${h_{2d}^z(x,y) =  - \frac{{{\nabla _t} \times e_{2d}^t(x,y) + i{k_0}\bar{\bm \mu} _r^{zt} \cdot h_{2d}^t(x,y)}}{{i{k_0}\bar{\bm \mu} _r^{zz}}}}$, leading to the following equation given by 
\begin{equation}\label{geq}
\bar{\bm{L}}{\bm \phi}_i=\beta_i \bar{\bm{B}}{\bm \phi}_i,
\end{equation}
where \begin{widetext}
\begin{equation}\label{L}
\bar{\bm{L}} = \left( {\begin{array}{*{20}{c}}
{  {D_1}\frac{{{D_2}}}{{{k_0}\bar{\bm \mu} _r^{zz}}} - {k_0}\bar{ \bm \epsilon} _r^{tt}{\rm{ + }}{k_0}\bar{ \bm \epsilon} _r^{tz}\frac{{\bar{ \bm \epsilon} _r^{zt}}}{{\bar{ \bm \epsilon} _r^{zz}}}}&{ - {k_0}\bar \chi _{eh}^{rtt} +i {D_1}\frac{{\bar{\bm \mu} _r^{zt}}}{{\bar{\bm \mu} _r^{zz}}} +i \bar{ \bm \epsilon} _r^{tz}\frac{{{D_2}}}{{\bar{ \bm \epsilon} _r^{zz}}}}\\
{ i {D_1}\frac{{\bar{ \bm \epsilon} _r^{zt}}}{{\bar{ \bm \epsilon} _r^{zz}}} +i \bar{\bm \mu} _r^{tz}\frac{{{D_2}}}{{\bar{\bm \mu} _r^{zz}}} + {k_0}\bar \chi _{he}^{rtt}}&{{-D_1}\frac{{{D_2}}}{{{k_0}\bar{ \bm \epsilon} _r^{zz}}} + {k_0}\bar{\bm \mu} _r^{tt} - {k_0}\bar{\bm \mu} _r^{tz}\frac{{\bar{\bm \mu} _r^{zt}}}{{\bar{\bm \mu} _r^{zz}}}}
\end{array}} \right),
\end{equation}
\end{widetext}
$\bar{\bm{B}} = \left[ {\begin{array}{*{20}{c}}
  0&0&0&{ - 1} \\ 
  0&0&1&0 \\ 
  0&{ - 1}&0&0 \\ 
  1&0&0&0 
\end{array}} \right]$, ${D_1} =\left( {\begin{array}{*{20}{c}}
  \frac{\partial }{\partial y} \\ 
  - \frac{\partial }{\partial x} \\ 
\end{array}} \right)$, ${D_2} =\left( {\begin{array}{*{20}{c}}
  { - \frac{\partial }{{\partial y}}}&{\frac{\partial }{{\partial x}}} 
\end{array}} \right)$, $\bm{ \phi}=[e_x,e_y,h_x,h_y]^T$, and $T$ is the transpose operation. \Equation{geq} completely determines  the series of waveguide modes labeled by $i$, with the propagation constant $\beta_i=n_{eff^i}k_0$, where $n_{eff^i}$ is the effective modal index.

It is a trivial step to reformulate the generalized eigenvalue form in \Eq{geq} to the  Hamiltonian form given by \Eq{ham}, i.e., $\mathcal{H}=\bar{\bm{B}}^{-1}\bar{\bm{L}}$, as derived in our previous work \cite{xzf_arxiv}. The waveguide Hamiltonian $\mathcal{H}$ gives a concise and complete description of the waveguide\cite{Sakurai1994}, but involves  complicated relations among the elements that are hard to interpret. In contrast, the generalized eigenvalue problem $(\bar{\bm{L}},\bar{\bm{B}})$ shows a much more transparent relation among the elements in $(\bar{\bm{L}},\bar{\bm{B}})$. It is easy to identify that $\bar{\bm{B}}$ is an  antisymmetric matrix, i.e., $\bar{\bm{B}}=-\bar{\bm{B}}^T$,  and some  blocks in $\bar{\bm{L}}$  are symmetric matrices. 
Considering the complementary  properties between $(\bar{\bm{L}},\bar{\bm{B}})$  and $\mathcal{H}$, we will use  both of them  where appropriate to describe the waveguide system. The rationale behind  will become  clear in the following discussions of this paper.

\subsection{ Left eigenvector ${\bm \psi}_j^T$ and scalar inner product}
% In this subsection,we define the left and right generalized eigenvector of $(L,\bar{\bm{B}})$ system, and simply discuss the B-orthogonal relation of them. The orthogonality is the fundamental of GCMT. And the left generalized eigenvector has close relationship with adjoint field. Real inner product is a general inner product, can be applied to various symmetry or non-symmetry operator. When the operator is hermitian, it becomes complex inner product. Real inner product is the fundamental of GCMT, which is used to transform the operator to a matrix to calculate. And it reveals the independent of the left and right vector.

%Generalized eigenvalue problem $L{\bm {\bm \phi}}=\beta \bar{\bm{B}} {\bm {\bm \phi}}$ can be easily transformed into a simple eigenvalue problem $\bar{\bm{B}}^{-1} L {\bm {\bm \phi}}=\beta {\bm {\bm \phi}}$ by left multiplying both sides by the inverse of $\bar{\bm{B}}$ when $det (\bar{\bm{B}})\neq 0$. So we can define the right generalized eigenvector ${\bm {\bm \phi}}$ in generalized eigenvalue problem \eqref{geq} like right eigenvector ${\bm {\bm \phi}}$ in eigenvalue problem\eqref{equ:eigenvalue problem}. 

Equation \eqref{geq} defines  the right eigenvectors of $(\bar{\bm{L}},\bar{\bm{B}})$, i.e., the expansion function ${\bm { \phi}}_i$ from the complete mode set $\left[{\bm { \phi}}_i\right]$. Similarly,  the left  eigenvector ${\bm \psi}$ is given by,
\begin{equation}\label{equ:left vector}
 \bar{\bm{L}}^a  {\bm \psi}_j= \beta \bar{\bm{B}}^a {\bm \psi}_j 
\end{equation}
where ${\bm \psi}_j$ has the same dimension as ${\bm { \phi}}_i$ but spans the complete mode set  $\left[{\bm \psi}_j\right]$ associated with the test function space. Following Rumsey \cite{Rumsey1954}, the inner product between ${ {\bm \phi}}_i$ and ${\bm \psi}_j$ is defined as following throughout the paper, 
\begin{equation}\label{equ:real inner product}
\left\langle {{\bm \psi}_j,{\bm { \phi}}_i} \right\rangle =\iint {{\bm \psi}_j^{T} {\bm { \phi}}_i dV},
\end{equation}
where ${\bm { \phi}}_i$/${\bm \psi}_j$ is the expansion/test  function. In the following part of this paper, the inner product defined in \Eq{equ:real inner product} is referred as the first type inner product throughout the paper.  We further introduce the second type inner product defined as follows,
\begin{equation}\label{B:inner product}
\left\langle {{\bm \psi}_j,{ {\bm \phi}}_i} \right\rangle_{\sigma} =\iint {{\bm \psi}_j^{T} \sigma { {\bm \phi}}_i dV},
\end{equation}
where the metric tensor $\sigma$ equals to $\bar{\bm{B}}$. The second type inner product can be easily reduced to the first type if the metric tensor $\sigma$ is an identity matrix, thus the first inner is coined as the identity inner product, while the second inner product is coined as the B-inner product. It will be shown that the  identity inner product is designed to work with the generalized eigenvalue description defined by $(\bar{\bm{L}},\bar{\bm{B}})$, while the B-inner product  works together with the Hamiltonian description of waveguide defined by $\mathcal{H}$.

\subsection{Adjoint waveguide system}
Provided the original waveguide described by $\mathcal{H} {\bm {\phi}}_i=\beta_i {\bm \phi}_i$, by definition the adjoint system $\mathcal{H}^a$ satisfies the following relation,
\beq\label{HamAdjoint}
\left\langle {{\bm \psi}_j, \mathcal{H}{\bm \phi}_i} \right\rangle_{\bar{\bm{B}}} +\left\langle {\mathcal{H}^a {\bm \psi}_j, {\bm \phi}_i}, \right\rangle_{\bar{\bm{B}}} =0,
\eeq 
for any ${\bm \phi}_i$ and ${\bm \psi}_j$ under the B-inner product. Considering the  equivalent generalized eigenvalue problem of  adjoint system $\mathcal{H}^a$  given by 
\begin{equation}\label{adjoint equation}
\bar{\bm{L}}^a {\bm \psi}_j = \beta_j \bar{\bm{B}} ^a {\bm \psi}_j,
\end{equation}
it is straight forward to prove that the adjoint relation given by \Eq{HamAdjoint} can be translated into the following relation between $(\bar{\bm{L}},\bar{\bm{B}})$ and $(\bar{\bm{L}}^a,\bar{\bm{B}}^a)$,
\begin{equation}\label{adjointconditionL}
\langle {\bm \psi}_j,\bar{\bm{L}} {\bm \phi}_i \rangle = \langle \bar{\bm{L}}^a {\bm \psi}_j, {\bm \phi}_i \rangle,
\end{equation}
where  $\bar{\bm{B}}^a=-\bar{\bm{B}}$. \Equation{adjointconditionL} indicates that the adjoint  relation between $(\bar{\bm{L}},\bar{\bm{B}})$ and $(\bar{\bm{L}}^a,\bar{\bm{B}}^a)$ can be simplified into the adjoint relation between $\bar{\bm{L}}$ and $\bar{\bm{L}}^a$. At present, the adjoint operator $\bar{\bm{L}}^a$ is in an abstract form, and the concrete form of $\bar{\bm{L}}^a$ will be given in the next section.

% It might be enlightening  to compare the transpose of \Eq{equ:left vector} with  \Eq{adjoint equation}, and one may expect that  the original system $(\bar{\bm{L}},\bar{\bm{B}})$ and adjoint system $(\bar{\bm{L}}^a,\bar{\bm{B}}^a)$ are related by  $\bar{\bm{B}}^a=\bar{\bm{B}}^T$,  $\bar{\bm{L}}^a= \bar{\bm{L}}^T$.  The examination in the next section shows $\bar{\bm{L}}^a \neq \bar{\bm{L}}^T$. 

% One indeed can prove that the projected forms, i.e., matrix representation under the projection onto the dual mode sets of $\left[{\bm { \phi}}_i\right]$  and $\left[{\bm { \psi}}_j\right]$, of  $(\bar{\bm{L}},\bar{\bm{B}})$ and $(\bar{\bm{L}}^a,\bar{\bm{B}}^a)$   share the same eigenvalues $\beta$, see details in Appendix A.  

% Importantly one can further prove that $(\bar{\bm{L}}^a,\bar{\bm{B}}^a)$ and $(\bar{\bm{L}}^T,\bar{\bm{B}}^T)$  share the same  eigenstates ${\bm \psi}_j$, meaning that by definition  $(\bar{\bm{L}}^a,\bar{\bm{B}}^a)$ and $(\bar{\bm{L}}^T,\bar{\bm{B}}^T)$ are exactly equivalent to each other under the scalar inner product, see details in Appendix A. 

\section{Reciprocal waveguides}

% waveguide reciprocity and orthogonal relation

% Symmetric relation between the

% coupled mode equation in

In section II., we have discussed that  the adjoint waveguide  system  defined by $(\bar{\bm{L}}^a, -\bar{\bm{B}})$ can be related to the original waveguide system via \Eq{adjointconditionL}, wherein  $\bar{\bm{L}}^a$ is implicit. In this section, we continue to discuss the adjoint waveguide system of  $(\bar{\bm{L}}, \bar{\bm{B}})$ and give explicit form of $\bar{\bm{L}}^a$  by reciprocity.  We further study the orthogonal relation between the modes from the two complete mode sets $\left[{\bm \phi}_i\right]$ and $\left[{\bm \psi}_i\right]$.

\subsection{Waveguide reciprocity and orthogonal relation}

Considering the  operator $\bar{\bm{L}}$ associated with the original waveguide problem  described by \Eq{L}, the adjoint operator $\bar{\bm{L}}^a$ can be readily given by the following form,
\begin{widetext}
\begin{equation}\label{adjoint operator}
{\bar{\bm{L}}^a} = \left( {\begin{array}{*{20}{c}}
{  {D_1}\frac{{{D_2}}}{{{k_0}\bar{\bm \mu }_{r}^{a,zz}}} - {k_0}\bar{\bm \varepsilon }_{r}^{a,tt}{\rm{ + }}{k_0}\bar{\bm \varepsilon }_{r}^{a,tz}\frac{{\bar{\bm \varepsilon }_{r,a}^{zt}}}{{\bar{\bm \varepsilon }_{r}^{a,zz}}}}&{ - {k_0}\bar{\bm \chi }_{r,eh}^{a,tt} +i {D_1}\frac{{\bar{\bm \mu }_{r}^{a,zt}}}{{\bar{\bm \mu }_{r}^{a,zz}}} +i \bar{\bm \varepsilon }_{r}^{a,tz}\frac{{{D_2}}}{{\bar{\bm \varepsilon }_{r}^{a,zz}}}}\\
{ i {D_1}\frac{{\bar{\bm \varepsilon }_{r}^{a,zt}}}{{\bar{\bm \varepsilon }_{r}^{a,zz}}} +i \bar{\bm \mu }_r^{a,tz}\frac{{{D_2}}}{{\bar{\bm \mu }_{r}^{a,zz}}} + {k_0}\bar{\bm \chi }_{r,he}^{a,tt}}&{{-D_1}\frac{{{D_2}}}{{{k_0}\bar{\bm \varepsilon }_{r}^{a,zz}}} + {k_0}\bar{\bm \mu }_{r}^{a,tt} - {k_0}\bar{\bm \mu }_{r}^{a,tz}\frac{{\bar{\bm \mu }_{r}^{a,zt}}}{{\bar{\bm \mu }_{r}^{a,zz}}}}
\end{array}} \right),
\end{equation}
\end{widetext}
where the material tensors  $\bar{\bm\varepsilon}_r^{a}$, $\bar{\bm\mu}_r^{a}$, $\bar{\bm \chi }_{r,he}^{a,tt}$, $\bar{\bm \chi }_{r,eh}^{a,tt}$ are to be  determined. According to the requirement imposed by the adjoint relation, i.e., \Eq{adjointconditionL}, it is straight forward to identify the material tensors of the adjoint waveguides, i.e., $\bar{\bm{\varepsilon}} _{r} ^a=\bar{\bm{\varepsilon}} _{r} ^T$,
$\bar{\bm\mu} _{r}^a =\bar{\bm\mu} _{r} ^T $, 
$\bar{\bm \chi }_{r,he}^{a,tt} =  - \left( \bar{\bm \chi }_{r,he}^{tt} \right)^T$, and $\bar{\bm \chi }_{r,eh}^{a,tt} =  - \left(\bar {\bm\chi} {_{r,he}^{tt}} \right)^T$. As for the reciprocal waveguide, the material tensors fulfill the reciprocity conditions, which require $\bar{\bm\varepsilon}_{r}=\left(\bar{\bm\varepsilon}_{r}\right)^T$, $\bar{\bm\mu}_r=\left( \bar{\bm\mu}_r \right)^T$, $\bar{\bm \chi }_{r,he}=-\left(\bar{\bm \chi }_{r,eh} \right)^T$. As such, it is trivial to find out that the following relation holds,
\begin{equation} \label{waveguidereciprocity}
\left\langle {{\bm \psi}_j ,\bar{\bm{L}}{\bm \phi}_i } \right\rangle  = \left\langle {{\bar{\bm{L}}}{\bm \psi}_j ,{\bm \phi}_i } \right\rangle, 
\end{equation}
which reveals that $\bar{\bm{L}}$  is self-adjoint under the scalar inner product defined by \Eq{equ:real inner product}. In contrast to the established equivalence of adjointness of the matrix form $\bm H$ of the Maxwell's equations in Xu's work \cite{Xu2015} and Lorentz reciprocity, where $\bm H$ is 3D operator, the operator $\bar{\bm{L}}$ is a 2D differential operator. Interestingly, the self-adjointness of the operator $\bar{\bm{L}}$ is also a necessary and sufficient condition to material reciprocity. Thus, we refer \Eq{waveguidereciprocity} as the waveguide reciprocity.

The waveguide reciprocity for reciprocal waveguides is one of the important  results of this paper, which we shall discuss  the relevance in relation with the generalized coupled mode formalism in depth later. To this end, we first examine the orthogonal relations of the waveguide modes. Combining the  original equation \Eq{geq} and the adjoint equation \Eq{adjoint equation} in the form of $\iint \left[{\bm \psi}_j\cdot\eqref{geq}-\eqref{adjoint equation}\cdot{\bm \phi}_i \right] dxdy$, it is straight forward to  derive the following equation \citep{Collin1960,McIsaac1991,Villeneuve1959},
\begin{equation}\label{equ:Lorentz reciprocity in original and adjoint field}
\left\langle {{\bm \psi}_j ,\bar{\bm{L}}{\bm \phi}_i } \right\rangle  - \left\langle {{\bar{\bm{L}}^a}{\bm \psi}_j ,{\bm \phi}_i } \right\rangle  = \left( {\beta_i  - {\beta_j}} \right)\left\langle {{\bm \psi}_j ,\bar{\bm{B}}{\bm \phi}_i } \right\rangle.
\end{equation}
\Equation{equ:Lorentz reciprocity in original and adjoint field} is essentially corresponding to the Lorentz reciprocity, where the source terms are set to be 0. Since we are interested in the reciprocal waveguide, the waveguide reciprocity requires that the term on the left hand vanishes, leading to the  orthogonal relation between ${\bm \phi}_i$  and ${\bm \psi}_j$ as follows,
\begin{equation}\label{equ:orthogonal}
(\beta_i-\beta_j)\iint{{\bm \psi}_j^T \bar{\bm{B}}{\bm \phi}_i \,dxdy}=0.
\end{equation}
For $\beta_i \neq \beta_j$, the term $\iint{  {\bm \psi}_j^T \bar{\bm{B}}{\bm \phi}_i \,dxdy}$ has to vanish, i.e, $\iint{  {\bm \psi}_j^T \bar{\bm{B}}{\bm \phi}_i \,dxdy}=0$. With proper normalizaion, the formula \Eq{equ:orthogonal} can be reformulated as follows,
\beq\label{kron}
\left\langle {{\bm \psi}_j,{\bm \phi}_i} \right\rangle_{\bar{\bm{B}}}=\delta_{ij},
\eeq 
where $\delta_{ij}$ is  Kronecker $\delta$ function. \Equation{kron} is referred as the B-orthogonal relation\cite{Silvestre2000} between  the original field ${\bm \phi}_i$ and the adjoint field ${\bm \psi}_j$ in this paper. By writing out all the components of ${\bm \phi}_i$ and ${\bm \psi}_j$ explicitly,  one find out that  $\iint{{\bm \psi}_j^T \bar{\bm{B}}{\bm \phi}_i\, dxdy}$ equals to  $\iint{\left( {e_i^t \times h_j^t - e_j^t \times h_i^t} \right)_z \,dxdy}$, which has the physical meaning of unconjugated form of Poynting vector along propagation direction \cite{Snyder}.

\subsection{Symmetric modal relations between ${\bm \phi}_i$ and  ${\bm \psi}_i$ in reciprocial waveguide}

\begin{table}
\begin{ruledtabular}
\caption{\label{4modeRelation}Symmetric relation of original field and adjoint field in the reciprocity waveguides with $\beta_i>0$}
\begin{tabular}{c|cc}
 &  mode with $\beta_i$  & mode with  $-\beta_i$  \\ \hline
  $(\bar{\bm{L}},\bar{\bm{B}})$  & $\left[\beta_i,\bm \phi_i\right]$ &     $\left[-\beta_i,\bm \psi_i\right]$  \\
  $(\bar{\bm{L}}^a,\bar{\bm{B}}^a)$ & $\left[\beta_i,\bm \psi_i\right]$   & $\left[-\beta_i,\bm \phi_i\right]$  \\
\end{tabular}
\end{ruledtabular}
\end{table}

By definition of adjoint operator, one can prove that the two complementary waveguide modes  described by  $(L^a,B^a)$ and $(L,B)$ share the same eigenvalues $\beta$ regardless self-adjointness of $L$, which is called the \textit{same-$\beta$} argument onwards in our paper, see proof in the Appendix A. Explicitly, the \textit{same-$\beta$} argument is  described by 
\begin{subequations}\label{SameBeta}
\begin{align}
\bar{\bm{L}} {\bm \phi}_i =\beta_i \bar{\bm{B}} {\bm \phi}_i, \label{SameBetaA} \\ 
\bar{\bm{L}}^a {\bm \psi}_i =\beta_i \bar{\bm{B}}^a {\bm \psi}_i, \label{SameBetaB}
\end{align}
\end{subequations}
where the original field $\phi_i$  and the adjoint field $\psi_i$ share the same $\beta_i$. As for reciprocal waveguide $L=L^a$ and anti-symmetry $\bar{\bm{B}}^a=-\bar{\bm{B}}$, \Eq{SameBetaB} can be reformulated as 
\beq\label{SameBetaC}
\bar{\bm{L}} {\bm \psi}_i =-\beta_i \bar{\bm{B}} {\bm \psi}_i.
\eeq
In comparison with \Eq{SameBetaA}, \Eq{SameBetaC} gives a different eigen solution $\left[-\beta_i, \bm \psi_i\right]$, apart from the eigen solution $\left[\beta_i, \bm\phi_i\right]$, both of which are directly associated  with $(L,B)$. The two different eigen-solutions share the same absolute value of $\beta_i$ but with different sign\cite{ZhuMultigridFiniteElementMethods}, which is called the \textit{paring-$\beta$} argument in this paper. Notably, the \textit{paring-$\beta$} argument also applies to the adjoint operator $(L^a,B^a)$ in reciprocal waveguides, meaning that if $\left[\beta_i, \bm\psi_i\right]$ is an eigen solution to $(L^a,B^a)$, see \Eq{SameBetaB}, there must be  a different solution $\left[-\beta_i, \bm\phi_i\right]$ which fulfills  $L^a\bm\phi_i=-\beta_i  B^a \bm\phi_i$. 

Evident from the aforementioned discussions, there are two modes ($\left[\beta_i, \bm\phi_i\right]$ and $\left[-\beta_i, \bm\psi_i\right]$) related with the original waveguide defined by $(L,B)$ for a given $\beta_i$. Importantly, the two modal fields ($\left[-\beta_i, \bm\phi_i\right]$ and $\left[\beta_i, \bm\psi_i\right]$)   are also the solutions to  $(L^a,B^a)$, but with the flipped sign of $\beta_i$. The   two inferred adjoint eigen solutions associated with $(L^a,B^a)$ from the known solutions of  $(L,B)$ can be a great help to construct  coupled mode equations in a generic bianisotropic waveguides, which will be discussed in the following sections.

The symmetric modal relations, dictated by the \textit{same-$\beta$} argument and the \textit{paring-$\beta$} argument, are summarized compactly in \Table{4modeRelation} for reciprocal waveguides. The established symmetric relation in \Table{4modeRelation}  is largely derived from the mathematical terms, which can also be interpreted with physical meanings. For a given $\beta_i$, i.e., $\beta_i>0$ corresponding to the forward propagating mode, the paring modes given by  $\left[\beta_i,\bm \phi_i\right]$ and $\left[-\beta_i,\bm \psi_i\right]$  are  essentially the forward backward propagating waveguide modes. From \textit{paring-$\beta$} argument,  one immediately realizes that the forward and backward propagating modes share the same absolute value of $\beta$. However, for a generic anisotropic/bianisotropic waveguide, there is no sign to show that the mode profiles of  forward and backward propagating modes, i.e., $\bm \phi_i$ and $\bm \psi_i$,  are necessarily the same or can be correlated. 

In a few special cases, the forward and backward propagating modes can indeed be transformed into each other via additional symmetries\cite{xzf_arxiv}, as tabulated in \Table{3Relation}. Once the symmetry relation  between the forward and backward propagating modes is known, one is able to use the forward propagating modes as the complete mode set to expand the field of the perturbed waveguide, and the backward propagating modes as the test function to construct the couple mode equation. In Hau's CMT \cite{HausJLT1987}, the time reversal operator $\mathcal{T}$ is used to infer $\bm \psi_i$ from $\bm \phi_i$, while the revised CMT in Xu's work \cite{Xu2015} takes the advantage of the chiral symmetry to infer $\bm \psi_i$ from $\bm \phi_i$. The single mode set used in  either Haus's CMT or Xu's work  is the one spanned by $\bm \phi_i$, i.e., $\left[\bm \phi_i\right]$. In generic bianisotropic waveguide, the symmetric relation between the forward and backward propagating modes vanishes, thus there is no simple way to deduce the $\bm \psi_i$ from $\bm \phi_i$. In this scenario, one need to combine   $\bm \psi_i$ and  $\bm \phi_i$ to form the complete mode set to construct CMT, which will be discussed in the next section.

\begin{table}
\begin{ruledtabular}
\caption{\label{3Relation}Symmetric relation of original field and adjoint field in the reciprocity waveguides}
\begin{tabular}{c|cc}
  Type  & Operator &  Symmetry relation  \\ \hline
  Chiral symmetry   &  $\bm\sigma$ &   ${\bm\psi}_i=\bm\sigma{\bm\phi}_i$  \\
  Time reverse symmetry & $\mathcal{T}$  & ${\bm\psi}_i=\mathcal{T}{\bm\phi}_i$\\
  Parity symmetry & $\mathcal{P}$  & ${\bm\psi}_i=\mathcal{P}{\bm\phi}_i$\\
\end{tabular}
\end{ruledtabular}
\end{table}

\subsection{Generalized coupled mode formalism by perturbation}

By perturbation, we construct the generalized coupled mode equations that treat the forward propagating  modes and the backward propagating modes on the same footing. Under a small perturbation on $\bar{\bm{L}} ^{\#}$, i.e., $\bar{\bm{L}} ^{\#} =\bar{\bm{L}}  + \Delta \bar{\bm{L}} $, the eigen-modes $\Phi$ associated with the perturbed  waveguide  $(\bar{\bm{L}} ^{\#},\bar{\bm{B}})$  can be expanded as by the eigen-modes $\phi_i$ of the unperturbed waveguide $(\bar{\bm{L}},\bar{\bm{B}})$. Explicitly, the perturbed waveguide mode is given by $\Phi=\sum a_i {\bm \phi}_i$, where $a_i$ are the coefficients to be determined. Under the small perturbation, the self-adjointness of operator $\bar{\bm{L}}$ in  reciprocity waveguide still holds, thus we can rewrite the \Eq{waveguidereciprocity} as follows, 
\begin{equation} \label{pertubation}
\left\langle {{\bm \psi_j} ,\bar{\bm{L}}  ^{\#}  {\bm \Phi} } \right\rangle  = \left\langle {{\bar{\bm{L}}}{\bm \psi}_j ,{\bm \Phi} } \right\rangle,
\end{equation}
where $\psi_j$ is the adjoint  modes associated with waveguide defined by $(\bar{\bm{L}},-\bar{\bm{B}})$. Importantly, the complete mode set in $[\psi_j]$ can be deduced from the known solutions of $\phi_i$, as evident from the \textit{same-$\beta$ argument} and the \textit{paring-$\beta$ argument} in our previous discussions.

In the following, it is trivial to plug the $\Phi=\sum a_i {\bm \phi}_i$ into \Eq{pertubation} and test \Eq{pertubation} against all possible $\psi_j$  in mode set $[\psi_j]$. With the assistance of the mode orthogonal relation, i.e., \Eq{equ:orthogonal}, one can derive the matrix form of our generalized couple mode equations as follows,
\begin{equation}\label{equ:CMT}
\Sigma {a_j}\left[ {{k_{ij} +b_{ij}} - i\left( {{\beta _i} - \beta } \right){p_{ij}}} \right] = 0
\end{equation}
where the boundary term ${b_{ij}}$ is given by  $
{b_{ij}} = -\frac{i}{{\bar{ \bm \epsilon} _r^{zz}}}\oint {[ {\varepsilon _r^{zt}e_i^th_j^t - {{\left( {\bar{ \bm \epsilon} _r^{tz}} \right)}^T}e_j^th_i^t} ] \cdot dl} 
$, and the normalzied term ${p_{ij}}$ is $
{p_{ij}} = -i\iint z \cdot ( {e_j^t \times h_i^t - e_i^t \times h_j^t} )\,dxdy$. The coupling coefficient $k_{ij}$ contains three  terms, i.e., $k_{ij}=k^1_{ij}+k^2_{ij}+k^3_{ij}$.  The first term  $k^1_{ij}$ is conventional perturbation contributed from transverse electric field, ${k^1_{ij}} = \iint { - \frac{{{k_0}}}{{\bar{ \bm \epsilon} _r^{zz}}}{e_j^{t}} \cdot \left( {\Delta \bar{ \bm \epsilon} _r^{tz}\bar{ \bm \epsilon} _r^{zt} + \bar{ \bm \epsilon} _r^{tz}\Delta \bar{ \bm \epsilon} _r^{zt} + \Delta \bar{ \bm \epsilon} _r^{tz}\Delta \bar{ \bm \epsilon} _r^{zt}} \right)e_i^tdxdy}$. The second term $k^2_{ij}$ stems from magnetoelectric coupling, i.e., $k^2_{ij}=i{k_0} \iint ( e{{_j^t}\cdot}\Delta \bar \chi _{r,eh}^{tt}h_i^t + e{{_i^t}}\cdot\Delta \bar \chi _{r,eh}^{tt}h_j^t )\,dxdy$, which could be particularly useful to study the mode hybridization in bianisotropic waveguides. The last term $k^3_{ij}$ is contributed from the coupling between the  transverse  field components and longitude  field components, i.e., $k^3_{ij}=\iint {}\frac{{k_0}}{{\bar{ \bm \varepsilon} _r^{zz}}}[ e{{_i^t}}\cdot\Delta \varepsilon _r^{tz}\left( {\bar{ \bm \varepsilon} _r^{zt}e_j^t +  \varepsilon _r^{zz}e_j^z} \right)+ e{{_j^t}}\cdot\Delta \varepsilon {{_r^{tz}}}\left( {\bar{ \bm \varepsilon} _r^{zt}e_i^t + \varepsilon _r^{zz}e_i^z} \right) ]\,dx dy$.

Close examination shows that  $b_{ij}$ in \Eq{equ:CMT} vanishes, thus the generalized  coupled mode equation of \Eq{equ:CMT} can be reduced as 
\begin{equation}\label{equ:final}
\sum\limits_j {{a_j}\left( {{\beta _j}{p_{ij}} - i{k_{ij}}} \right)}  = \sum\limits_j {{a_j}\beta {p_{ij}}}, 
\end{equation}
where $\beta$ and the  $a_j$ are propagation constant and modal expansion coefficients. The present coupled mode equation  resembles the same  matrix form as in Haus's CMT, as well as that in our own work. However, it is worthy to emphasize that  both the forward and backward propagating modes are included in the mode expansion set in our formula \Eq{equ:final}. We will refer \Eq{equ:final} as GCMF,  in order  to be distinct from previous coupled mode equations, typically from Haus and ours \cite{HausJLT1987,Xu2015}.   

As an example, we give the explicit matrix form of GCMF for two modes hybridization, in which the definite relation between the forward and backward propagating modes does not exist during the perturbation. In this regard, there are four modes, i.e., two forward propagating modes (${\bm \phi}_1$ ,${\bm \phi}_2$) and two backward propagating modes (${\bm \psi}_{1}$, ${\bm \psi}_{2}$), spanning the complete expansion mode set. Meanwhile, reciprocity guarantees that the  mode set associated with the adjoint waveguide system  $(\bar{\bm{L}}^a, \bar{\bm{B}}^a)$ are the same as that of   $(\bar{\bm{L}},\bar{\bm{B}})$  due to the fact that $\bar{\bm{L}}=\bar{\bm{L}}$, and $\bar{\bm{B}}^a=-\bar{\bm{B}}$. Therefore, the test function can be  simply chosen from the four modes ${\bm \phi}_1$ ,${\bm \phi}_2$,  ${\bm \psi}_{1}$, ${\bm \psi}_{2}$. The  eigen modes of  perturbed waveguide is given by $\Phi=a_1\phi_1+a_2\phi_2+a_3\psi_1+a_4\psi_2$, in which the coefficients $a_j$ and the propagation constant $\beta$ can be determined by the following coupled mode equation, 
\begin{widetext}
\begin{equation}\label{the matrix form of GCMF}
\left( {\begin{array}{*{20}{c}}
  {{\beta _1}{p_{11}} - i{k_{11}}}&{{\beta _2}{p_{12}} - i{k_{12}}}&{{\beta _3}{p_{13}} - i{k_{13}}}&{{\beta _4}{p_{14}} - i{k_{14}}} \\ 
  {{\beta _1}{p_{21}} - i{k_{21}}}&{{\beta _2}{p_{22}} - i{k_{22}}}&{{\beta _3}{p_{23}} - i{k_{23}}}&{{\beta _4}{p_{24}} - i{k_{24}}} \\ 
  {{\beta _1}{p_{31}} - i{k_{31}}}&{{\beta _2}{p_{32}} - i{k_{32}}}&{{\beta _3}{p_{33}} - i{k_{33}}}&{{\beta _4}{p_{34}} - i{k_{34}}} \\ 
  {{\beta _1}{p_{41}} - i{k_{41}}}&{{\beta _2}{p_{42}} - i{k_{42}}}&{{\beta _3}{p_{43}} - i{k_{43}}}&{{\beta _4}{p_{44}} - i{k_{44}}} 
\end{array}} \right)\left( \begin{gathered}
  {a_1} \hfill \\
  {a_2} \hfill \\
  {a_3} \hfill \\
  {a_4} \hfill \\ 
\end{gathered}  \right) = \beta \left( {\begin{array}{*{20}{c}}
  {{p_{11}}}&{{p_{12}}}&{{p_{13}}}&{{p_{14}}} \\ 
  {{p_{21}}}&{{p_{22}}}&{{p_{23}}}&{{p_{24}}} \\ 
  {{p_{31}}}&{{p_{32}}}&{{p_{33}}}&{{p_{34}}} \\ 
  {{p_{41}}}&{{p_{42}}}&{{p_{43}}}&{{p_{44}}} 
\end{array}} \right)\left( \begin{gathered}
  {a_1} \hfill \\
  {a_2} \hfill \\
  {a_3} \hfill \\
  {a_4} \hfill \\ 
\end{gathered}  \right).
\end{equation}
\end{widetext}

\section{Results and discussions}
In this section, we apply the generalized coupled mode equation, i.e. \eqref{equ:final},  to study the modal coupling in anisotropic and bianisotropic waveguide that may contain gain and losses. 

\begin{figure*}[]
\centering
\includegraphics[width=0.7\textwidth]{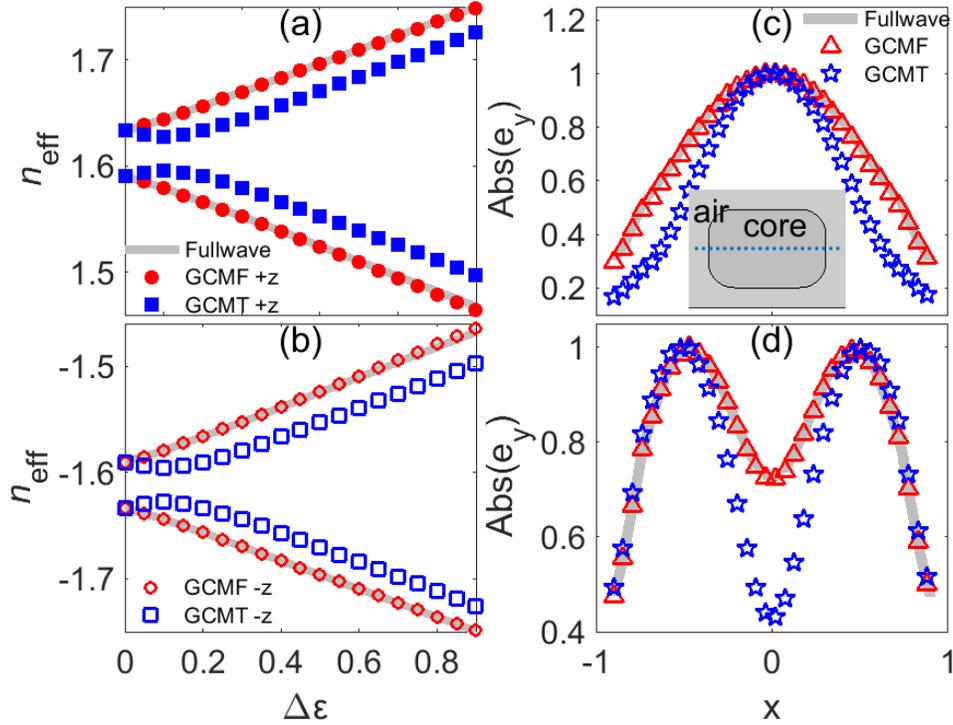}
\caption{ (a/b) Effective modal  indices  $n_{eff}$ as function of increased magnitude of anisotropy  $\Delta \varepsilon$ for forward/backward propagating modes, calculated from GCMF (red solid circles/ red open
circles), GCMT \cite{Xu2015} (blue solid circles/blue open circles), and fullwave simulation using finite element method (gray line). In (c/d), $|e_y|$ along the dotted line  shown in the inset for the mode  $\phi_1$  and the mode $\phi_2$ at $\Delta \varepsilon=0.2$ is shown. The field plots are calculated from  GCMF (red triangles), GCMT (blue stars) and fullwave simulation (gray line). The inset in (c) shows  the cross section of the waveguide with width $a=0.375 \lambda_0$, height $b=0.25\lambda_0$, where vacuum wavelength $\lambda_0$. The dotted line shows the spatial coordinates, where $e_y$  is plotted.} 
\label{fig1}
\end{figure*}

\subsection{Anisotropic waveguide}
In the first example, we study the anisotropic waveguide, see waveguide structure in the inset in \Fig{fig1}. The dielectric tensor  of the  the waveguide core is anisotropic, i.e.,  $\bar{\bm\varepsilon}_r=\left(\begin{array}{*{18}{c}}
  {10}&0&{0.1} \\ 
  0&{10}&0 \\ 
  {0.1}&0&5 
\end{array} \right)+\Delta \varepsilon \left(\begin{array}{*{18}{c}}
  {0}&0&{1} \\ 
  0&{0}&0 \\ 
  {1}&0&0 
\end{array} \right)$， where $\Delta \varepsilon$ is the strength of the perturbation accounting for the magnitude of anisotropy in the off-diagonal terms in   $\bar{\bm\varepsilon}_r$. The particular selection of anisotropic perturbation will be discussed later.  The surrounding medium is air.  \Figure{fig1} shows the comparison of dispersion diagrams obtained from three independent approaches, i.e., GCMF (this paper), GCMT \cite{Xu2015} and full finite element modeling using commercial software package, COMSOL MULTIPHYSICS \cite{comsol}, as  $\Delta\varepsilon$ varies. In this waveguide, there are two  forward propagating modes, and two  backward propagating  modes. The modal indices $n_{eff}$ are  further separated as  $\Delta\varepsilon$ increases, as shown  in \Fig{fig1} (a-b). The modal indices between forward and backward propagating modes are symmetric with respect to $n_{eff}=0$ due to the \textit{paring-$\beta$  argument} in reciprocal waveguides, which also holds in this example.     

% These three mirror-symmetric lines reveal different implications: The Fullwave line is calculated from finite element method so it proves that pairing-$\beta$ argument that the forward eigenvalue is opposite of backward's is correct. The GCMF's mirror line is calculated in a natural way (we don't add any condition to make the the line must be mirror), so it could prove GCMF is correct on some level. For GCMT, the mirror line is coming from that we set the backward line is opposite of forward line. 

As can be seen from \Fig{fig1}, the dispersion plots obtained from GCMF (red circles) show excellent agreement with the numerical results from the fullwave simulations, but with a large discrepancy with that obtained from GCMT\cite{Xu2015}. To further examine discrepancy, we plot $|e_y|$ of the mode $\Phi_1$/$\Phi_2$ along the cutline sketched in the inset in \Fig{fig1} (c/d) at $\Delta\varepsilon=0.2$, wherein red triangles from GCMF, blue pentagrams from GCMT, and  the gray line from fullwave simulation. The spatial dependent field again shows that the results from GCMF and fullwave simulation match well, with a large discrepancy from that obtained from GCMT.

The discrepancy stems from the broken link between the forward and backward propagating modes, i.e., the chiral symmetry, which has been used implicitly to construct coupled mode equations in Xu's work. Due to the presence of the off-diagonal terms in the dielectric tensor, the  transverse components  $\bm e_t$ of the electric field of the waveguide mode are coupled to the longitudinal component  $e_z$. Notably, the  intrinsic spin-momentum locking of waveguide modes \cite{Bliokh} gives rise to $\beta$-dependent relation between transverse components $\bm e_t$ and longitudinal component $e_z$. For instance, ${{\mathbf{e}}^z} = \frac{{\left( {{\nabla _t} \cdot {{{\mathbf{\bar \varepsilon }}}_r}^{tt}} \right){{\mathbf{e}}^t} + {\nabla _t} \cdot {{{\mathbf{\bar \varepsilon }}}_r}^{tz}{{\mathbf{e}}^z}}}{{i\beta {\varepsilon _r}^{zz}}} - \frac{{\left( {{{{\mathbf{\bar \varepsilon }}}_r}^{zt}} \right){{\mathbf{e}}^t}}}{{{\varepsilon _r}^{zz}}}$, ${{\mathbf{h}}^z} = \frac{{{{\mathbf{h}}^t}}}{{i\beta }}$. Supposing that the in-plane electric fields of forward and backward propagation modes are identical, the longitude component is reduced to ${{\bf{e}}^z} = \frac{{\left( {{\nabla _t} \cdot {{{\bf{\bar \varepsilon }}}_r}^{tt}} \right){{\bf{e}}^t}}}{{i\beta {\varepsilon _r}^{zz}}}$ when $\bar {\bm\varepsilon}_r^{zt}=0$ and $\bar {\bm\varepsilon}_r^{tz}=0$. Considering the opposite propagation constant in forward and backward modes, the longitude components of forward and backward propagation modes share the same absolute value but with different sign which is known as chiral symmetry. However, the longitude components of electric fields of forward and backward mode are not the opposite number when $\bar {\bm\varepsilon}_r^{zt}\ne 0$ or $\bar {\bm\varepsilon}_r^{tz}\ne 0$, that is, the chiral symmetry relation is lost.

\begin{figure}[]
\centering
\includegraphics[width=0.5\textwidth]{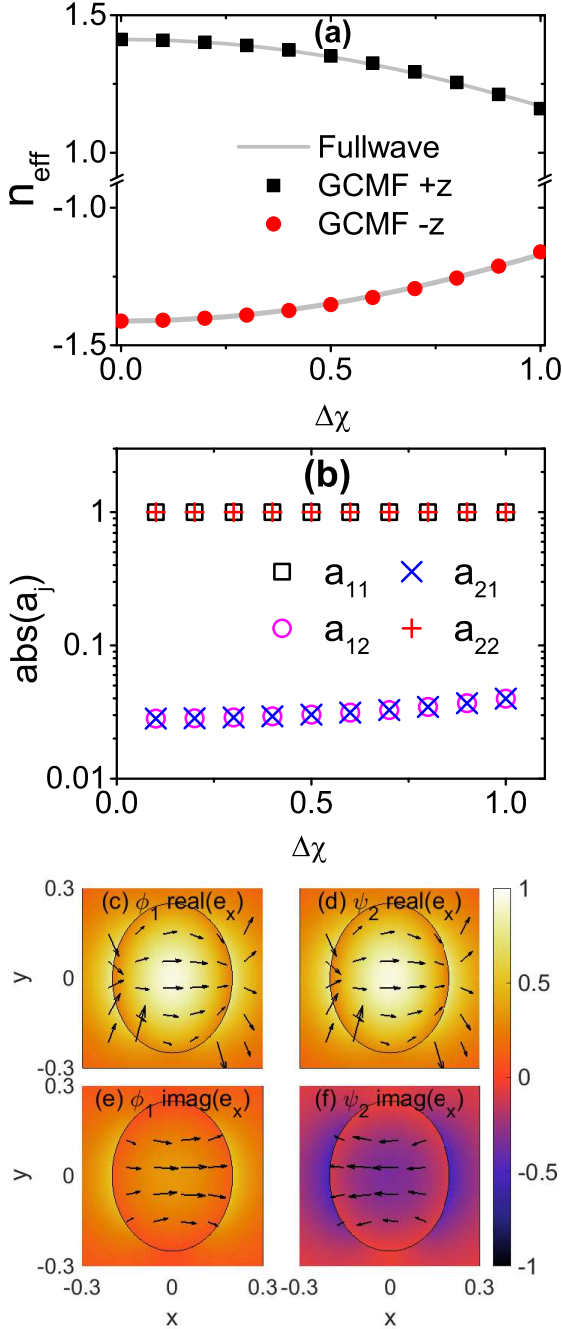}
\caption{(a) Effective modal indices $n_{eff}$ as function of the magnetoelectric coupling ($\Delta\chi$)  that accounts for the bianisotropy. The cross section of the waveguide is elliptical,  with long (short) axis $a=0.25 \lambda_0$ ($b=0.2 \lambda_0$).  (a) shows the $n_{eff}$ of forward/backward propagating x-polarized mode denoted by black/red solid circles. The gray solid lines are calculated from fullwave simulations. (b) shows modal coefficients of the hybridized mode upon a small perturbation,  as $\Delta\chi$ varies. The real part of x-component of forward/backward normalized electric field is shown in (c)/(d), and the vector plots of the real part of in-plane electric field are also shown in (c)/(d) indicated by the arrows, the length of which is proportional to the magnitude of the vector field. (e-f) show the imaginary part of field, and the other setting is same to (c-d). In (c-f),  the field profiles are obtained from COMSOL at $\Delta\chi=0.4$.}
\label{fig2}
\end{figure}

\subsection{Bianisotropic waveguide}
In the second example, we consider a bianisotropic meta-material waveguides, which can be realized by aligning electrically small split-ring resonators  along particular direction, see details in Xu's paper\cite{bian}.  In this reciprocal  waveguide,  the geometric configuration is identical to conventional waveguide with high material index in the core surrounded by air. In addition, the core layer contains bianisotropy, which is described by \Eq{DB} with $\varepsilon_r=4$,   $\mu_r=1$, and $\bar{\bm \chi} _{r,eh}=-\left( \bar{\bm \chi} _{r,eh} \right)^T=\left(\begin{array}{*{18}{c}}
  0& i\Delta\chi  & 0 \\ 
  0 & 0 & 0 \\ 
  0 & 0  & 0 
\end{array} \right)$.  The purpose of studying this typical bianisotropic waveguide is to illustrate the relevance of including the  backward propagating modes in the modal expansion set to obtain the correct modal hybridization, which will become clear shortly. 

In this bianisotropic waveguide, there are two forward propagating modes, i.e., the x-polarized mode and y-polarized mode. In the presence of $\Delta \chi$,   the x-polarized mode will be  altered significantly.  In contrast, the y-polarization mode will not  be affected at all, which is not shown here. We apply GCMF to study the mode hybridization  in \Fig{fig2} (a), which shows the effective modal index as function of $\Delta \chi$.  The red symbols representing  modal indices calculated by GCMF match well with the gray line obtained by fullwave simulations. Apparently, the \textit{paring-$\beta$ argument} still applies. And the modal index decreases for larger  $\Delta \chi$, see detailed explanation in Xu's work\cite{Xu_bianisotropic2015}.

% Moreover, it is easy to find out two facts from \Fig{fig2} (a): (1) the paring-$\beta$ argument still applies; (2) the modal index decreases for larger $\Delta \chi$. 

Though the  x-polarized mode and y-polarized mode are completely decoupled as $\Delta \chi$ varies, the forward propagating x-polarized mode and the backward propagating mode are coupled  together. In the implementation of GCMF, the y-polarized mode has been excluded, thus the complete modal set in constructing GCMF contains two modes, which are the forward and backward propagating x-polarized mode, i.e., $\phi_1$, $\psi_2$. Under perturbation, the two  hybridized modes can be given by $\Phi_1=a_{11}\phi_1+ a_{12}\psi_2$, $\Psi_2=a_{21}\phi_1+ a_{22}\psi_2$, where the cross modal coefficient $a_{12}$ ($a_{21}$) refers to the contribution from backward (forward) propagating mode $\psi_2$ ($\phi_1$) to the newly hybridized forward (backward) propagating mode  $\Phi_1$ ($\Psi_2$). \Fig{fig2} (b) shows the normalized modal coefficients obtained from GCMF in the modal hybridization  between the forward and backward propagating x-polarized mode.  The modal coefficients $a_{ij}$ are  complex numbers, here only the absolute values of  $a_{ij}$ are shown.  Apparently,  the diagonal terms  $a_{11}$ and $a_{22}$  are dominating, while  the off-diagonal terms $a_{12}$ and $a_{21}$, up to ten percentage, are not negligible, see \Fig{fig2} (b). The non-negligible value  of 
 $a_{12}$ indeed confirms our expectation that the backward propagating mode will contribute to the forward propagating mode in the modal hybridization under perturbation of  $\Delta \chi$, and vice versa.

% As the $\bar{\bm\chi_{r,eh}}$ $\bar{\bm\chi_{r,he}}$ is imaginary number, so this example is a time reverse hermitian system. The effective mode index $n_{eff}$ is a real number, which satisfy the calculation result in FIG. 2(a) and 2(b).  From Xiong' work, one system satisfy the chiral symmetry $\bm\sigma \mathcal{H} \bm\sigma^{-1}=-\mathcal{H} $ when $\bar{\bm\chi}_{r,eh}=0$, $\bar{\bm\chi}_{r,he}=0$ and the field has the relation $\bm\psi_i=\bm\sigma\bm\phi_i$. Due to the existence of $\Delta\chi$, this model is not a chiral symmetry system. FIG.2(e)(f)(g)(h) shows the real part and imaginary part of $E_y$ field calculated from GCMF at $\Delta\chi=0.4$. From the figure, it can be seen that $\bm\phi_1 \ne \bm\sigma\bm\psi_1$. So using chiral  operation to get backward field using GCMT may be wrong in bianisotropic waveguide. 

The $e_x$ component of the mode is shown in FIG. 2 (c-f) in order to illustrate that GCMT could not apply to this example in detail. In FIG. 2. (c)/(d), the real part of x-component of forward/backward electric field and the vector plots of the real part of in-plane electric field are shown. And in (e)/(f) the imaginary part of field is plotted. And these fields are all obtained from COMSOL. From the picture, it is clear that the in-plane vector of forward field $\phi_1$ and the backward field $\psi_2$ are same while their imaginary part are opposite. It means the forward field and backward field are not chiral symmetry because chiral symmetry means their real part and imaginary are both identical. So still using the chiral symmetry will get the false backward field due to the opposite the imaginary part. 

As investigated in Xiong's work \cite{xzf_arxiv}, and from FIG. 2. (c-f), the forward and backward propagating modes are time reversal pairs. Once the coupled mode equation is implemented in the complex inner product, the expansion modal set  shall contain only one mode, which is enough due to the time reversal symmetry. In the next section, we will continue to study the waveguide  containing gain, losses, as well as bianisotropy, in which both the chiral symmetry and the time reversal symmetry are broken. In those generic bianisotropic waveguide, the only corrected way to construct the coupled mode equations is to simultaneously include the forward and backward propagating modes in the modal expansion set.

\begin{figure*}[]
\centering
\includegraphics[width=0.7\textwidth]{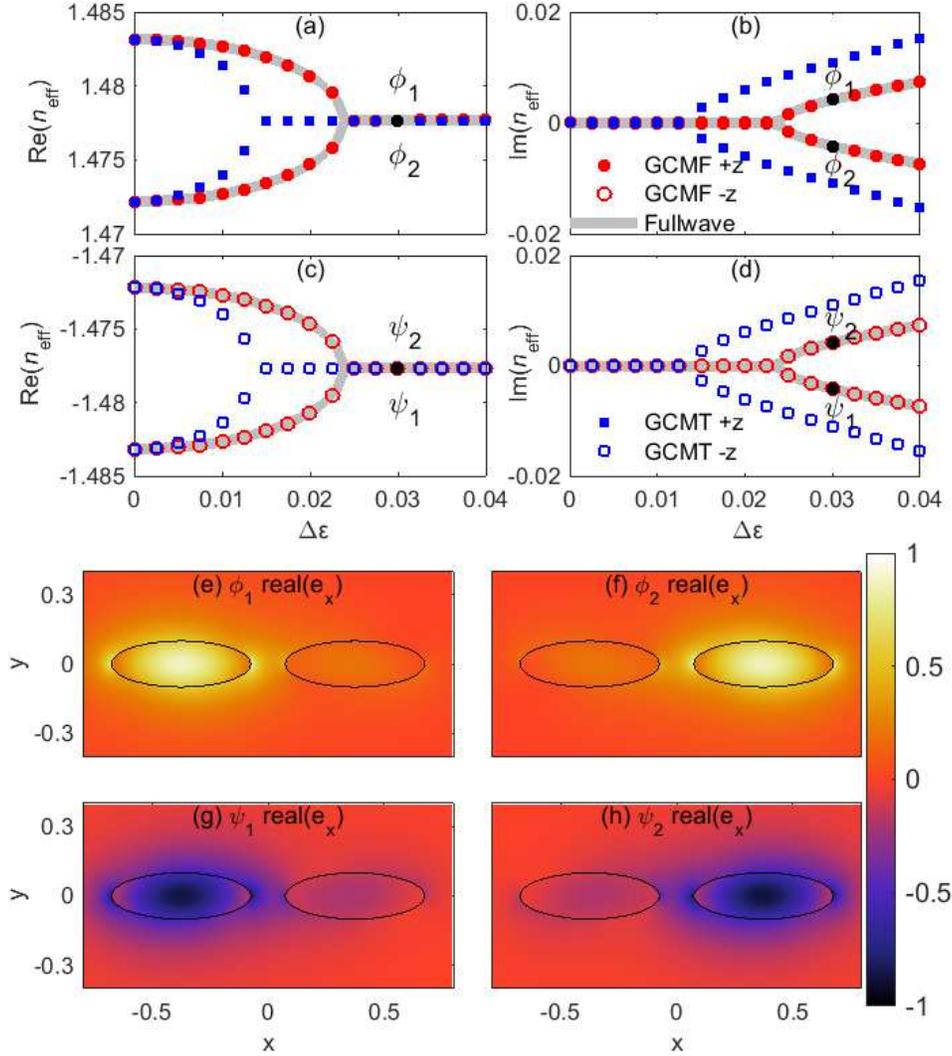}
\caption{Effective mode indices $n_{eff}$ versus perturbation $\Delta\varepsilon$ using GCMF, GCMT and fullwave simulation. The structure is a double elliptical waveguide with width long axis $a=0.3\lambda_0$, short axis $b=0.1\lambda_0$ and the center distance $d=2.5a$. The red/blue/gray marker represents the result from GCMF/GCMT/Fullwave respectively. The (a-b)/(c-d) represent forward/backward mode $n_{eff}$ respectively. (a-c)/(b-d) represent real/imaginary part of $n_{eff}$ respectively. The normalized $Re(e_x)$ at $\Delta \varepsilon=0.03$ calculated by GCMF are shown in (e-h). (e)/(g) represent the forward/backward field of mode 1, and (f)/(h) represent the forward/backward field of mode 2. $\lambda_0$ is the vacuum wavelength}
\label{fig:non-symmetry structure}
\end{figure*}

\subsection{Bianisotropic waveguide with $\mathcal{PT}$ symmetric gain and losses}

The third example is a $\mathcal{PT}$ symmetry optical system with balanced gain and losses, as well as bianisotropy. The structure contains two identical single mode elliptical waveguides surrounded by the air cladding, see FIG.3. The material in left elliptical guide is $\varepsilon_r=4+i\Delta\varepsilon$, $\chi^{xx}_{r,eh}=-\chi^{xx}_{r,he}=1$, and in right elliptical guide is $\varepsilon_r=4-i\Delta\varepsilon$, $\chi^{xx}_{r,eh}=-\chi^{xx}_{r,he}=-1$. The magnetoelectric coupling terms $\chi$ in two ellipses are opposite in order to make this system have $\mathcal{PT}$-symmetry, i.e., $\mathcal{PT}\bar{\bm{L}}(\mathcal{PT})^{-1}=\bar{\bm{L}}$, see detail in Xiong's work\cite{xzf_arxiv}.

The unperturbed system has two supermodes, an odd mode and an even mode, which is hybridized by two identical mode from two single-mode waveguides. These two modes are coupled by the gain/loss perturbation $i\Delta\varepsilon$, which is a parameter to measure the non-Hermiticity of this system. As the parameter $\Delta\varepsilon$ increases, the system undergoes a phase transformation from a completely real spectrum into a complex spectrum, which is known as $\mathcal{PT}$-symmetry\cite{Klaiman2008,Chongyidong2017}. See in FIG.3(a)(b), the $n_{eff}$ of two modes obtained from GCMF are real numbers representing exact $\mathcal{PT}$-symmetry before the except point $\Delta \varepsilon=0.024$, while $n_{eff}$ of two modes appearing the imaginary part become the complex conjugate numbers after the except point. In FIG.3(c)(d), the backward mode shows exactly the same phenomenon.

As the $\mathcal{PT}$-symmetry occurs, two conjugated modes (two forward modes or two backward modes) may be obtained from one another under the subsequent $\mathcal{P}$ ($\bm r\rightarrow -\bm r$) and $\mathcal{T}$ (complex conjugate) operations. See FIG.3, $\bm \phi_1$ and $\bm \phi_2$ in (e)(f) are two forward modes which satisfy $\bm\phi_1(r)=\bm\phi_2^*(-r)$. And the backward modes $\bm \psi_1$ and $\bm \psi_2$ in (g)(h) satisfy the same symmetry, i.e. $\bm\psi_1(r)=\bm\psi_2^*(-r)$.

No matter whether the $\mathcal{PT}$-symmetry breaks or not, the chiral relation between the forward and backward propagation modes is broken due to the existence of $\bar{\bm\chi}_{r,eh}$ and $\bar{\bm\chi}_{r,he}$. The blue square symbols in FIG. 3(a)-3(d) representing effective mode indices given by GCMT show large discrepancy with the gray line calculated by COMSOL. It is clear that in this case GCMT fails to capture the major feature of bianisotropic waveguides. The red circular symbols in same figures calculated by our GCMF match excellently well with the results from fullwave simulations.

\section{Conclusion}
In conclusion,  we developed a generalized coupled mode formulation to study the mode hybridization in reciprocal waveguides, in which the anisotropy and the bianisotropy play an essential role. In our description, the waveguide problem is reformulated as a generalized eigenvalue problem as the original system, accompanied by the its adjoint generalized eigenvalue problem as its dual partner. The two complementary systems together define  the dual mode sets, which are needed in constructing the coupled mode theory. In reciprocal waveguides, we find out that the symmetry relations between the  dual mode sets are dictated by the \textit{same-$\beta$ argument }and the \textit{paring-$\beta$ argument}, which turns out be intimately related with the forward and backward propagating modes. Accordingly,  the generalized coupled mode theory that can be reduced to the existing coupled mode schemes, is realized  by treating the forward and backward propagating modes on the same footing in the modal expansion set. Importantly, the generalized coupled mode theory developed here handles the modal coupling in anisotropic and bianisotropic waveguides, where the existing coupled mode schemes fail. We illustrate the capability of our generalized coupled mode theory through three examples, i.e., anisotropic waveguides, bianisotropic waveguides, and bianisotropic waveguides with balanced gain and losses. The three examples unambiguously show the feasibility and the strength  of our theory in studying the mode hybridization in waveguides with broken link between the forward and backward propagating modes.

% Using real inner product, we establish the equivalence between adjoint relation and waveguide reciprocity condition, then we derive the mode orthogonal relation. For waveguide problem, when the Maxwell's equation is reduced to $2$D from $3$D, it becomes a generalized eigenvalue problem which is original system. The adjoint system is introduced from waveguide reciprocity condition to provide the adjoint field to test the original field.  Importantly, when the original and adjoint system is self-adjoint or reciprocity material, the adjoint field and original field are same but with opposite eigenvalues, so that the adjoint field can be transformed into backward field to test forward field. 

% Our theory provide a method for precisely calculating the waveguide mode propagation constant or cavity eigenfrequency after perturbation on material or shape and it might be useful to design device or integrated optics guides with bianisotropic material.

\section{Acknowledgment}
This work was supported in part by National Natural Science Foundation of China (Grant No. 61405066, 61405067, 61775063 and 61735006), National Key Research and Development Program of China (Grant No. 2017YFA0305200), and the Fundamental Research Funds for the Central Universities, HUST: 2017KFYXJJ027.

\appendix

\section{The proof of \textit{same-$\beta$} argument}
We consider the dual waveguide systems described by the two equations,
\begin{eqnarray}
\label{q1}
&\bar{\bm{L}} {\bm \phi}^1=\beta^1 \bar{\bm{B}} {\bm \phi}^1, \\
%\label{q2}
%&\bar{\bm{L}}^T {\bm {\bm \psi}}^2=\beta^2 \bar{\bm{B}}^T {\bm {\bm \psi}}^2 \\
\label{q3}
&\bar{\bm{L}}^a {\bm {\bm \psi}}^2=\beta^2 \bar{\bm{B}}^a {\bm {\bm \psi}}^2. 
\end{eqnarray}
In order to find the relation of eigenvalue $\beta^1$ and $\beta^2$, the bi-orthogonal basis $[{\bm \phi_i}]$, $[{\bm \phi_j}]$ is used as a basis to represent the above differential operator. The matrix elements of the operator in the $[{\bm \phi_i}]$, $[{\bm \phi_j}]$ basis are easily obtained by applying the standard Galerkin moment method as follows,
\begin{eqnarray}
\label{a}
&\bar{\bar{\bm L}}_{ij}= \iint{    {\bm {\bm \psi}} _i^T    \bar{\bm{L} }   {\bm \phi} _j\,dxdy          },\\
\label{c}
&\bar{\bar {\bm L}}^a_{ij} = \iint {{{\bm \phi} _i}^T{\bar{\bm{L}}^a}{{\bm \psi} _j}\,dxdy},\\
&\bar{ \bar {\bm{B}}}_{ij} = \iint {{{\bm \psi} _i}^T{\bar{\bm{B}}}{{\bm \phi} _j}\,dxdy},\\
&\bar{ \bar {\bm{B}}}^a_{ij} = \iint {{{\bm \phi} _i}^T{\bar{\bm{B}}^a}{{\bm \psi} _j}\,dxdy,}
\end{eqnarray}
where $\bar{ \bar {\bm L}}$, $\bar{ \bar {\bm L}}^a$, $\bar{ \bar {\bm B}}$, $\bar{ \bar {\bm B}}^a$ are matrix representation. Transposing adjoint relation \eqref{adjointconditionL}, one could derive 
\begin{equation}
\label{adjointMatrix}
\iint{{\bm \phi}_i^T \bar{\bm{L}}^T {\bm \psi}_j \,dxdy} = \iint{ {\bm \phi}_i^T \bar{\bm{L}}^a {\bm \psi}_j \,dxdy.} 
\end{equation}
Due to that $\bar { \bar {\bm{L}}}_{ij}$ is a scalar, one can transpose it without changing value, 
\begin{equation}
\bar { \bar {\bm{L}}}_{ij} = \iint {{{\bm \phi} _j}^T{\bar{\bm{L}}^T}{{\bm \psi} _i}\,dxdy},
\end{equation}
and subsequently we shall obtain
\begin{equation}
\label{Lji}
\bar { \bar {\bm{L}}}_{ji} = \iint {{{\bm \phi} _i}^T{\bar{\bm{L}}^T}{{\bm \psi} _j}\,dxdy}.
\end{equation}
Identifying the three equations, i.e., \eqref{Lji}, \eqref{q3} and \eqref{adjointMatrix}, one obtains
\begin{equation}
\bar { \bar {\bm{L}}}_{ji} =\bar{\bar {\bm L}}^a_{ij},
\end{equation}
which gives rise to the symmetric relation as, 
\begin{equation}
\bar {{\bar {\bm L}}}^a=\left(\bar {\bar {\bm L}}\right)^T.
\end{equation}
As for $\bar{\bm{B}}$, one easily derives
\begin{equation}
\bar { \bar {\bm{B}}}^a=\left(\bar { \bar {\bm{B}}}\right)^T=-\bar { \bar {\bm{B}}}.
\end{equation}
Using matrix form to reformulate the \Eq{q1}, \Eq{q3}, one immediately obtains
\begin{eqnarray}
\label{b1}
&\bar  {\bar {\bm L}} {\bm \phi}^1=\beta^1 \bar  {\bar {\bm{B}} }{\bm \phi}^1, \\
% \label{b2}
% &\overline {\bar {\bm L}^T} {\bm \psi}^2=\beta^2 \overline { \bar {\bm{B}}^T}{\bm \psi}^2 \\
\label{b3}
&\bar {\bar {\bm L}}^a {\bm \psi}^2=\beta^2 \bar  { \bar {\bm{B}}}^a {\bm \psi}^2. 
\end{eqnarray}
The  eigenvalue equations of \Eq{b1} and \Eq{b3} are $|\bar {\bar {\bm L}}-\beta^1\bar  { \bar {\bm B}} |=0$ and  $|(\bar {\bar {\bm L}})^T+\beta^2 \bar { \bar {\bm B}}|=0$, which are essentially the same via a transpose operation. Therefore we derive  the \textit{same-$\beta$ argument}, i.e., $
\beta^1=\beta^2$, which states that 
the  original and adjoint waveguide  systems share the same propagation constants $\beta$.

\end{document}